\documentclass[aps,preprint,amsmath,amssymb]{revtex4}

\usepackage{graphicx}
\begin{document}

\title{Tests of flavor universality for neutrino-Z couplings in
future neutrino experiments}

\author{A. B. Balantekin}
\email[]{baha@physics.wisc.edu}
\affiliation{Department of Physics,
University of Wisconsin, Madison, WI 53706, USA}

\author{\.{I}. \c{S}ahin}
\email[]{isahin@wisc.edu} \email[]{isahin@science.ankara.edu.tr}
\affiliation{Department of Physics, University of Wisconsin,
Madison, WI 53706, USA}
 \affiliation{Department of
Physics, Faculty of Sciences, Ankara University, 06100 Tandogan,
Ankara, Turkey}

\author{B. \c{S}ahin}
\email[]{bsahin@wisc.edu} \email[]{dilec@science.ankara.edu.tr}
\affiliation{Department of Physics, University of Wisconsin,
Madison, WI 53706, USA}
 \affiliation{Department of
Physics, Faculty of Sciences, Ankara University, 06100 Tandogan,
Ankara, Turkey}

\begin{abstract}
We investigate the physics potential of NuSOnG experiment to
probe new physics contributions to $Z\nu\nu$ couplings
in muon-neutrino electron elastic and neutral-current deep-inelastic
scattering processes. We employ an effective Lagrangian approach and
do not a priori assume universality of the coupling of neutrinos to Z. We
obtain $95\%$ C.L. limits on possible universality violating
couplings.
\end{abstract}

\maketitle

\section{Introduction}
Standard model (SM) of particle physics has been extensively tested by
experiments at CERN, Fermilab Tevatron and elsewhere. These
experimental results confirm the $SU_L(2)\otimes
U_Y(1)$ gauge structure of the SM. Measurement of gauge boson
couplings to fermions provide us important data for the
determination of $SU_L(2)$ and $U_Y(1)$ couplings. The charged
lepton couplings to Z boson have been measured with a sensitivity of
$O(10^{-4})$ \cite{Yao:2006px}. However Z boson couplings to
individual neutrinos have not been tested with comparably good accuracy.
 For example, the experimental limits on $\nu_e$ and $\nu_\mu$
couplings to Z are approximately 100 times worse than $e$ and $\mu$
couplings to Z \cite{Yao:2006px}. Universality of the coupling of
neutrinos to Z is another assumption of the SM which has not been
tested with a good accuracy. This assumption simply states that
$\nu_e$, $\nu_\mu$ and $\nu_\tau$ couple with the same strength to Z
at the tree-level.

Many parameters of the SM  have been very precisely tested at,
for example, CERN
$e^{+} e^{-}$ collider LEP. At LEP, couplings of neutrinos to Z are
constrained by the invisible Z width which receives contributions
from all neutrino flavors. Therefore it is impossible to
discern possible universality violating neutrino Z couplings
from the LEP data. It is possible to constrain new physics
contributions to $Z\nu\nu$ that respect universality. Recent limits
on these contributions are \cite{Yao:2006px,Masso:2002vj}
\begin{eqnarray}
|\Delta_e+\Delta_\mu+\Delta_\tau|\leq 0.009
\end{eqnarray}
where the parameters $\Delta_e,\Delta_\mu$ and $\Delta_\tau$ describe
possible deviations from the SM coming from new physics. They modify
neutrino neutral current as \cite{Masso:2002vj}
\begin{eqnarray}
\label{smcurrent}
J_\mu^{NC}=\frac{1}{2}\sum_i[1+\Delta_i]\bar{\nu}_i\gamma_\mu \nu_i
\end{eqnarray}
These new physics contributions respect universality if the equality
$\Delta_e=\Delta_\mu=\Delta_\tau$ holds.

CHARM II
Collaboration obtained data on $\nu_\mu e\to \nu_\mu e$
scattering. These data together with LEP results place the limit
\cite{Masso:2002vj,Vilain:1994gx}
\begin{eqnarray}
|\Delta_\mu|\leq 0.037 .
\end{eqnarray}
Using the limits given in Eqs. (1) and (3) we equivalently have the limit
\begin{eqnarray}
|\Delta_e+\Delta_\tau|\leq 0.046
\end{eqnarray}

The universality of $\nu_e$ and $\nu_\mu$ coupling to the neutral
weak current has also been tested experimentally by the CHARM
Collaboration \cite{Dorenbosch:1986tb}.
The ratio of the coupling constants is
given by $g_{\nu_e \bar{\nu}_e}/g_{\nu_\mu
\bar{\nu}_\mu}=1.05^{+0.15}_{-0.18}$. From this ratio and previous
limits, the following bounds can be obtained
\begin{eqnarray}
-0.13\leq -\Delta_\mu+\Delta_e \leq 0.20\, ,
\,\,\,\,\,\,\,\,\,\,\,\,\,\, -0.167\leq \ \Delta_e \leq 0.237 .
\end{eqnarray}

The processes impacting only a single neutrino
flavor could violate neutrino flavor universality and therefore
provide more information about new physics probes on $Z\nu\nu$
couplings compared to invisible decay width experiments of Z boson.
Recently a new, high energy, high statistics neutrino scattering
experiment, called NuSOnG (Neutrino Scattering on Glass), has been
proposed \cite{Adams:2008cm}.
Such a "terascale" (with energies of 1 TeV and beyond)
experiment could offer unprecedented physics opportunities.
NuSOnG experiment uses a
Tevatron-based neutrino beam to study $\nu_\mu e^-
\to \nu_\mu e^-$ and $\nu_\mu e^- \to \nu_e \mu^-$ reactions as
well as neutral- and charged-current deep inelastic scattering with high
statistics.

In this paper we investigate the physics potential of this future
experiment to probe possible new physics contributions to
$Z\nu\nu$ couplings. To carry out a more general treatment we do not
assume universality of the coupling of neutrinos to Z.

\section{Effective Lagrangian approach to $Z\nu\nu$ couplings}

There is an extensive literature on non-standard interactions of neutrinos
\cite{Davidson:2003ha,Bell:2005kz,Mohapatra:2005wg,Mohapatra:2006gs,de Gouvea:2007xp,Perez:2008ha}.
New physics contributions to neutrino-$Z$ couplings can be investigated
in a model independent way by means of the effective Lagrangian
approach. Specifically we consider the $SU(2)_L\otimes U(1)_Y$ invariant
effective Lagrangian introduced in Ref. \cite{Buchmuller:1985jz}. Possible
deviations from the SM that may violate neutrino-Z coupling
universality are described by the following dimension-6 effective
operators:
\begin{eqnarray}
\label{eop1}
O_j=i(\phi^\dagger D_\mu \phi)(\bar \ell_j \gamma^\mu \ell_j)\\
O^\prime_j=i(\phi^\dagger D_\mu \vec{\tau} \phi)\cdot(\bar \ell_j
\gamma^\mu \vec{\tau} \ell_j)
\label{eop2}
\end{eqnarray}
where $\ell_j$ is the left-handed lepton doublet for flavor
$j=e,\mu$ or $\tau$; $\phi$ is the scalar doublet; and $D_\mu$ is the
covariant derivative, defined by
\begin{eqnarray}
D_\mu=\partial_\mu+i\,\frac{g}{2}\,
\vec{\tau}\cdot\vec{W}_\mu+i\,\frac{g^\prime}{2}\,YB_\mu .
\end{eqnarray}
Here $g$ and $g^\prime$ are the $SU(2)_L$ and $U(1)_Y$ gauge
couplings, $Y$ is the hypercharge and the gauge fields $W^{(i)}_\mu$
and $B_\mu$ sit in the $SU(2)_L$ triplet and $U(1)_Y$ singlet
representations, respectively.

The most general $SU(2)_L\otimes U(1)_Y$ invariant Lagrangian up to
dimension-6 operators, containing new physics contributions that may
violate universality of the neutrino Z coupling, is then given by
\begin{eqnarray}
\label{lag}
{\cal L}={\cal
L}_{SM}+\sum_{j=e,\mu,\tau}\frac{1}{\Lambda^2}(\alpha_j\,O_j+\alpha^\prime_j
O^\prime_j)
\end{eqnarray}
where, ${\cal L}_{SM}$ is the SM Lagrangian, $\Lambda$ is the energy
scale of new physics and $\alpha_j$, $\alpha^\prime_j$ are the
anomalous couplings. After symmetry breaking, Lagrangian in Eq. (\ref{lag})
reduces to \cite{Buchmuller:1985jz}
\begin{eqnarray}
{\cal L}^\prime=\frac{g}{\cos\theta_W}\,J^{NC}_\mu\,Z^\mu ,
\end{eqnarray}
with
\begin{eqnarray}
\label{current}
J^{NC}_\mu=\left[\frac{1}{2}+
\frac{v^2}{2\Lambda^2}(-\alpha_j+\alpha^\prime_j)\right]
\bar {{\nu}_j}_{L}\gamma_\mu
{{\nu}_j}_{L}+\left[-\frac{1}{2}+
\sin^2\theta_W-\frac{v^2}{2\Lambda^2}(\alpha_j+\alpha^\prime_j)\right]
\bar {{\ell}_j}_{L} \gamma_\mu {\ell_j}_{L}
\end{eqnarray}
In this effective current subscript "L" represents the left-handed leptons
and $v$ represents the vacuum expectation value of the scalar
field. (For definiteness, we take $v=246$ GeV and $\Lambda = 1$ TeV
in the calculations presented in this paper).

As can be seen from the current in Eq. (\ref{current}), the operators of Eq.
(\ref{eop1}) and (\ref{eop2}) modify not only the neutrino currents,
but also the left-handed charged lepton currents.
On the other hand right-handed charged lepton currents
are not modified. We show in the next section that this fact has important
consequences in $\nu_\mu e^- \to \nu_\mu e^-$ scattering.

Comparing currents (\ref{smcurrent}) and (\ref{current}) we express the
parameters $\Delta_j$ in terms of couplings $\alpha_j$ and $\alpha^\prime_j$:
\begin{eqnarray}
\Delta_j=\frac{v^2}{\Lambda^2}(-\alpha_j+\alpha^\prime_j)
\end{eqnarray}
We see that the parameters $\Delta_j$ actually consist of two independent
couplings which need to be constrained by the experiments.

\section{$\nu_{\mu}-e$ elastic and neutral current deep inelastic
scattering}

Muon-neutrino electron elastic scattering is described by a t-channel
Z exchange diagram.
As we have discussed in the previous section, not only the $\nu_\mu\nu_\mu Z$
vertex
but also the $e^-e^-Z$ vertex is modified by the effective Lagrangian.
The differential cross section is given by
\begin{eqnarray}
\label{cs}
\frac{d\sigma(\nu_\mu e^- \to \nu_\mu e^-)}{dy}=\frac{2G_F^2 m_e
E_\nu}{\pi}\left(1+\frac{v^2}{\Lambda^2}(-
\alpha_\mu+\alpha_\mu^\prime)\right)^2
\left[\eta^2+\epsilon_+^2(1-y)^2-\eta \epsilon_+\frac{m_e}{E_\nu}y\right]\\
\nonumber \\
y=\frac{E_e^\prime-m_e}{E_\nu},\,\,
\,\,\,\,\,\,\,\,\,\,\,\,\,\,\,\,\,\,\,\,\,\,\,\,\,\,\,\,\,\,\,\,\,\,
0\leq y\leq \frac{1}{1+\frac{m_e}{2E_\nu}}\,\,\,\,\,\,\,
\,\,\,\,\,\,\,\,\,\,\,\,\,\,\,\,\,\,\,\,\,\,\,\,\,
\,\,\,\,\,\,\,\,\,\,\,\,\,\,\,\,\,\,\,\,\,\,\,\,\,\,\,\,\,\,\,\,
\end{eqnarray}
where $E_\nu$ and $E_e^\prime$ are the initial neutrino and
final electron energies,
$m_e$ is the mass of the electron, $G_F$ is the Fermi constant, $\Lambda$
is the energy scale of new physics and $v$ is the vacuum expectation value
of the scalar Higgs field. The parameters $\eta$ and $\epsilon_+$ appearing
above are defined as
\begin{eqnarray}
\label{eta}
\eta =&&-\frac{1}{2}+\sin^2\theta_W
-\frac{v^2}{2\Lambda^2}(\alpha_e+\alpha_e^\prime), \\
\epsilon_+=&&\sin^2\theta_W .
\end{eqnarray}
We see from Eqs. (\ref{cs}) and (\ref{eta}) that contribution of
$\alpha_e$ to the cross section is equal to the contribution of
$\alpha_e^\prime$. It is then impossible to
distinguish $\alpha_e$
from $\alpha_e^\prime$ and therefore we only consider the
coupling $\alpha_e$ in our numerical calculations.
The couplings  $\alpha_\mu$ and $\alpha_\mu^\prime$ can be distinguished from
$\alpha_e$ and $\alpha_e^\prime$ with the help of polarization. For
left-handed
final state electrons only the term proportional to $\eta^2$ contributes to the
differential cross section. On the other hand for right-handed final
state electrons
only the term proportional to $\epsilon_+^2$ contributes. Therefore
right-handed
cross section isolates the couplings $(-\alpha_\mu+\alpha_\mu^\prime)$.
The interference
term proportional to $\eta \epsilon_+$ does not contribute if we neglect
the mass of final electron.

Neutrino magnetic dipole moment is very small in SM, but it may receive
contributions from new physics. With the neutrino magnetic moment
there is a
t-channel photon exchange diagram which contributes to the process
$\nu_\mu e^- \to \nu_\mu e^-$. This contribution increases the cross section
by
\cite{Marciano:2003eq,Marciano:1987pf,Kyuldjiev:1984kz,Vogel:1989iv,Balantekin:2006sw}
\begin{eqnarray}
\label{magmom}
\frac{\Delta d\sigma(\nu_\mu e^- \to \nu_\mu
e^-)}{dy}=\mu^{2}\frac{\pi\alpha^{2}}{m_{e}^{2}}\left(\frac{1}{y}-1\right),
\end{eqnarray}
where $\mu$ is the neutrino magnetic moment measured in units of
Bohr Magnetons.
Consistency of $\nu_\mu e$ cross sections with SM expectations
tightly constrains the neutrino magnetic moment \cite{Marciano:1987pf},
$\mu <10^{-9} \mu_B$. Therefore the contribution (\ref{magmom}) is
very little especially for high energy neutrinos due to y dependence
\cite{Marciano:2003eq}. For this reason we will neglect this photon exchange
contribution.

\begin{figure}
\includegraphics{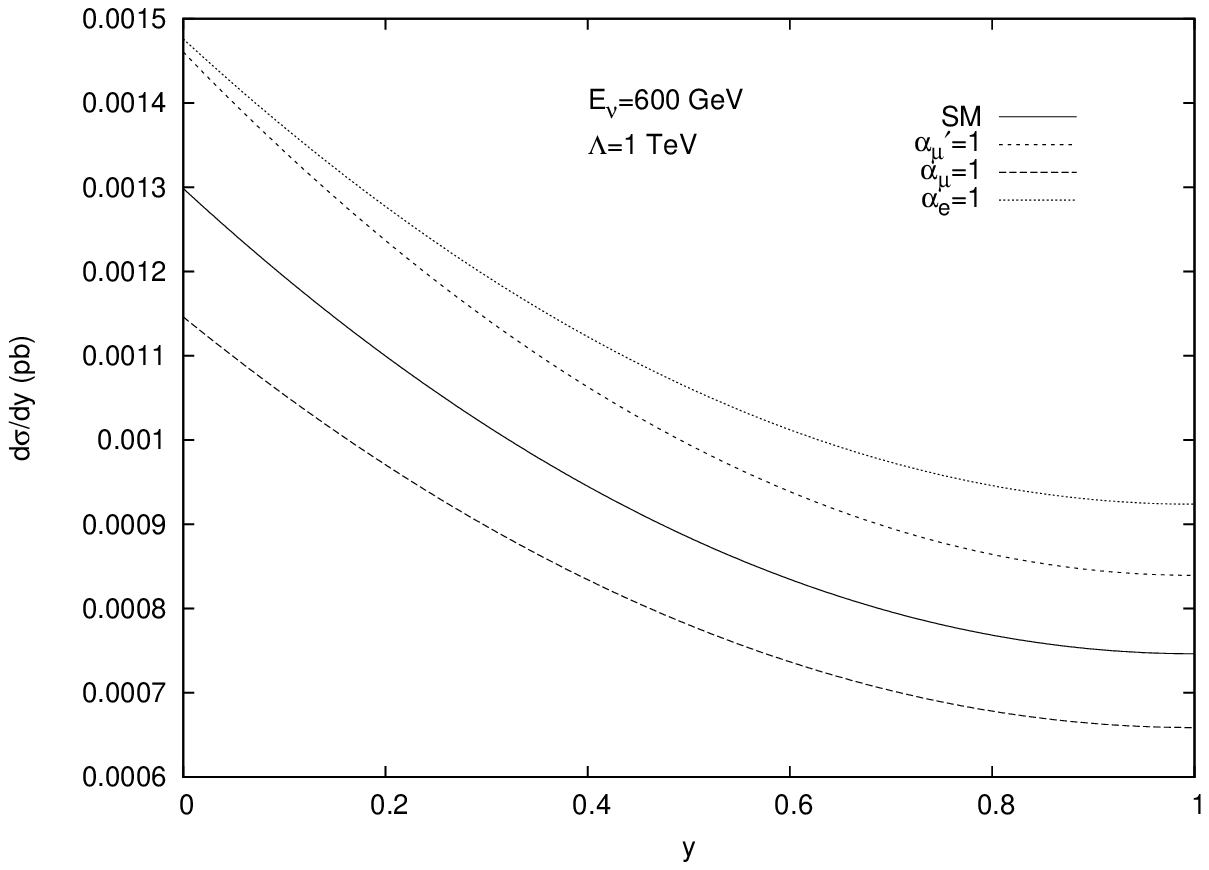}
\caption{Differential cross section as a function of y for
various values of the anomalous couplings. Only one of the
anomalous couplings is kept different from their SM value.\label{fig1}}
\end{figure}

In Fig. \ref{fig1} we plot the differential cross section as a function
of y for various values of the anomalous couplings. We see from this
figure that deviation of the differential cross section from its SM
value is larger for $\alpha_e=1$ as compared with $\alpha_\mu=1$ and
$\alpha_\mu^\prime=1$ cases. The shape of the curves for
$\alpha_\mu=1$ and $\alpha_\mu^\prime=1$ are exactly the same with the
SM curve.  But the behavior of $\alpha_e=1$ curve is slightly
different from the SM one. Its deviation from the SM increases as the
parameter y increases. Hence a terascale neutrino facility could in
principle probe physics that yields $\alpha_e \neq 0$.

\begin{figure}
\includegraphics{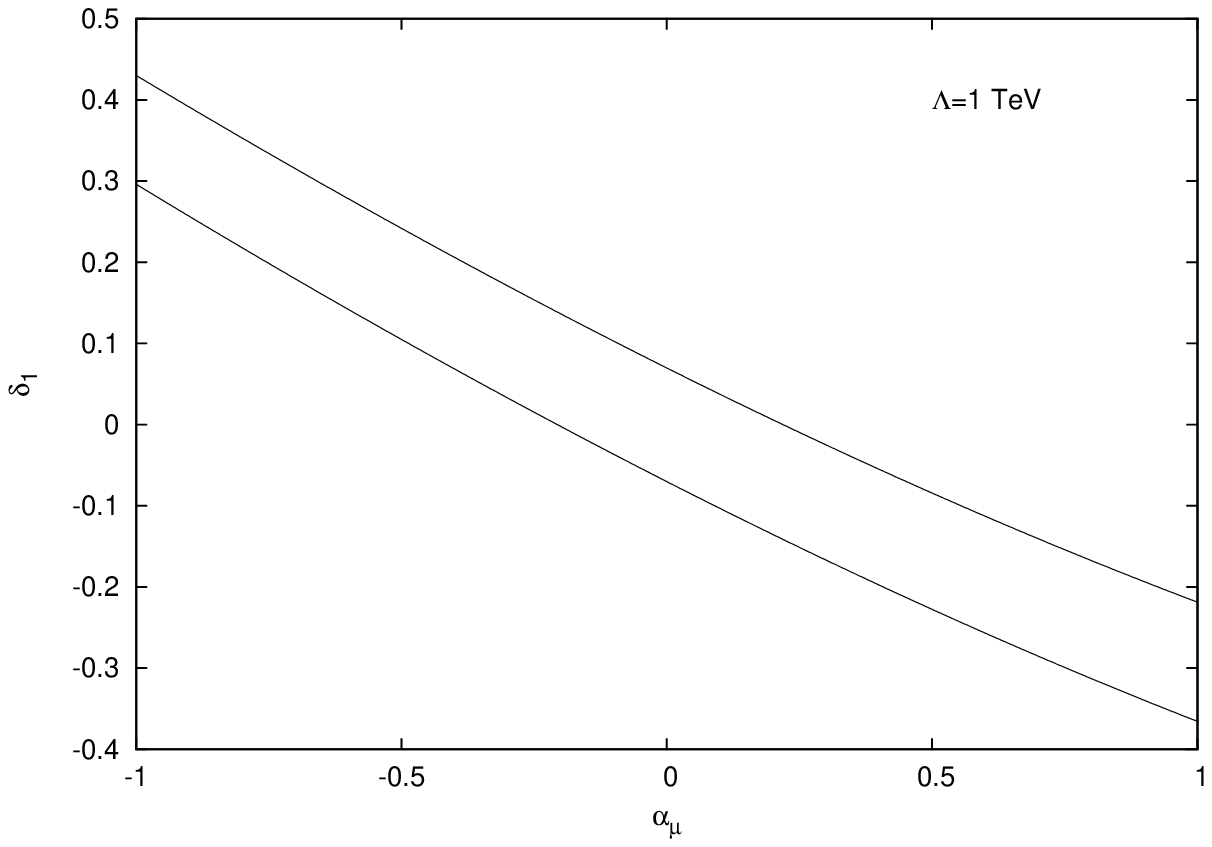}
\caption{$95\%$ C.L. sensitivity bound on the parameter space $\alpha_\mu - \delta_1$.
 Sensitivity bound is the area restricted by the lines.
 $\alpha_\mu^\prime$ is taken to be zero.
\label{fig2}}
\end{figure}

\begin{figure}
\includegraphics{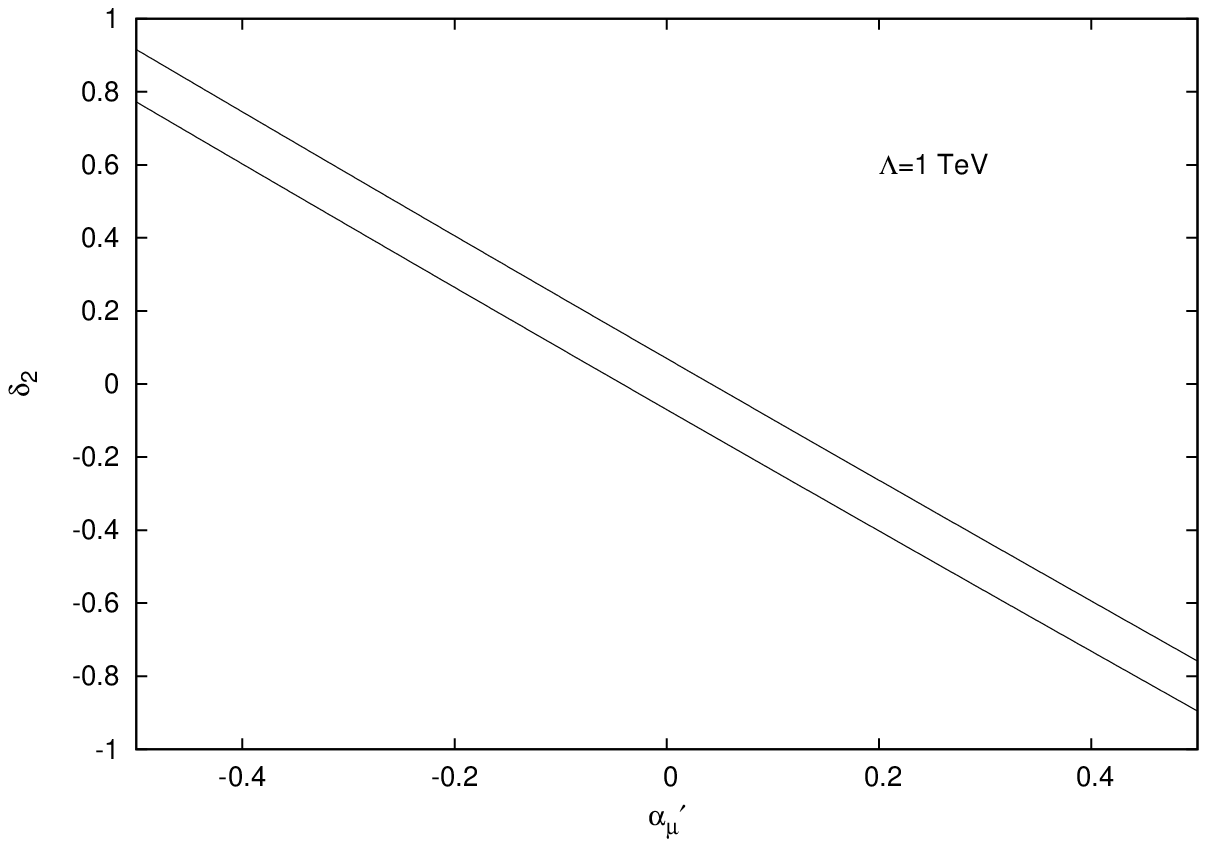}
\caption{$95\%$ C.L. sensitivity bound on the parameter space $\alpha_\mu^\prime - \delta_2$.
 Sensitivity bound is the area restricted by the lines.
 $\alpha_\mu$ is taken to be zero.
\label{fig3}}
\end{figure}

\begin{figure}
\includegraphics{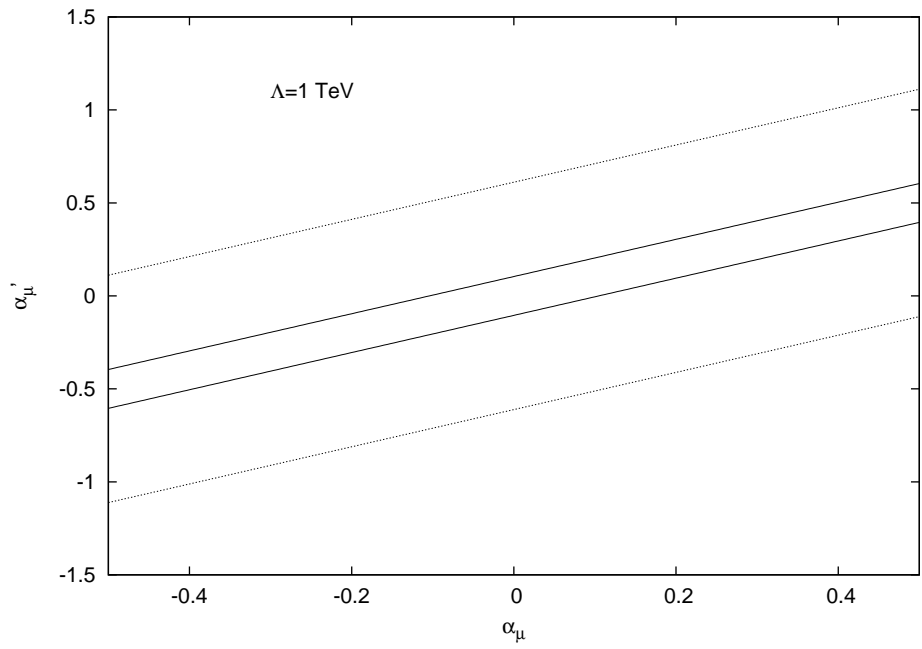}
\caption{The area bounded by the solid lines is $95\%$ C.L.
sensitivity bound on the parameter space $\alpha_\mu -
\alpha_\mu^\prime$. Dotted lines show the limits obtained from CHARM
II data. $\alpha_e$ is taken to be zero. \label{fig4}}
\end{figure}

Neglecting terms of order $\frac{m_e}{E_\nu}$, we obtain the
total cross section:
\begin{eqnarray}
\sigma(\nu_\mu e^- \to \nu_\mu e^-)&&=\frac{2 G_F^2 m_e E_\nu}{\pi}
\left(1+\frac{v^2}{\Lambda^2}(-\alpha_\mu+\alpha_\mu^\prime)\right)^2
\nonumber \\
&&\times\left[\frac{\sin^4\theta_W}{3}+\left(\frac{1}{2}-\sin^2\theta_W
+\frac{v^2}{2\Lambda^2}(\alpha_e+\alpha_e^\prime)\right)^2\right] .
\end{eqnarray}
We studied $95\%$ C.L. bounds using two-parameter $\chi^{2}$
analysis with a systematic error of the same order as the
statistical one. The $\chi^{2}$ function is given by,
\begin{eqnarray}
\chi^{2}=\left(\frac{\sigma_{SM}-\sigma_{AN}}{\sigma_{SM} \,\,
\delta_{exp}}\right)^{2}
\end{eqnarray}
where $\sigma_{AN}$ is the cross section containing new physics
effects and $\delta_{exp}=\sqrt{\delta_{stat}^2+\delta_{syst}^2}$.
$\delta_{stat}=\frac{1}{\sqrt{N}}$ is the statistical error and
$\delta_{syst}$ is the systematic error. Number of events is taken
to be $N=75000$ which is compatible with the number of events
studied at Ref. \cite{Adams:2008cm}.  We re-parametrize the
couplings as
\begin{eqnarray}
\alpha_e=\alpha_\mu+\delta_1=\alpha_\mu^\prime+\delta_2 .
\end{eqnarray}
Thus possible non-zero values of the couplings $\delta_1$ or
$\delta_2$ implies universality violation between interactions
$\nu_\mu\nu_\mu Z$ and $\nu_e\nu_e Z$: Any modification of the SM
$\nu_\mu\nu_\mu Z$ and $\nu_e\nu_e Z$ couplings that respect
universality is described by $\delta_1=\delta_2=0$ (or equivalently
by $\alpha_\mu=\alpha_\mu^\prime=\alpha_e$). In Fig. \ref{fig2},
\ref{fig3}, and \ref{fig4} we show $95\%$ C.L. allowed regions for
the parameter spaces $\alpha_\mu - \delta_1$, $\alpha_\mu^\prime -
\delta_2$ and  $\alpha_\mu - \alpha_\mu^\prime$.  In Fig. \ref{fig4}
we also show the limit (area bounded by dotted lines) obtained from
inequality (3).  We see from this figure that limit obtained from
the CHARM II data is approximately 6 times weaker than our limits.

NuSOnG experiment will also provide high statistics $\nu_\mu$ deep
inelastic scattering from the nuclei in glass.
The expected number of events
for $\nu_\mu$ neutral current deep inelastic scattering is
$190\times10^6$ \cite{Adams:2008cm}. In comparison
NuTeV had $1.62\times10^6$ deep
inelastic scattering events in neutrino mode \cite{zeller}.  Therefore
NuSOnG will provide two orders of magnitude more events.
Since quark couplings to Z are not modified by operators (6,7) hadron
tensor does not receive any contribution. It is defined by the
standard form  \cite{Blumlein:1996vs,Forte:2001ph}
\begin{eqnarray}
W_{\mu\nu}=\left(-g_{\mu\nu}+\frac{q_\mu
q_\nu}{q^2}\right)F_1(x,Q^2)+\frac{\hat{p}_\mu \hat{p}_\nu}{p\cdot
q}F_2(x,Q^2)-i\epsilon_{\mu\nu\alpha\beta} \frac{q^\alpha
p^\beta}{2p\cdot q} F_3(x,Q^2)
\end{eqnarray}
where $p_\mu$ is the nucleon momentum, $q_\mu$ is the momentum of Z
propagator, $Q^2=-q^2$, $x=\frac{Q^2}{2p\cdot q}$ and
\begin{eqnarray}
\hat{p}_\mu\equiv p_\mu-\frac{p\cdot q}{q^2}q_\mu \nonumber
\end{eqnarray}
The structure functions are defined as follows
\cite{Forte:2001zu,Forte:2002ip}
\begin{eqnarray}
F_1=&&\frac{1}{2}\sum_i (g_V^2+g_A^2)_i(q_i+\bar{q}_i) \\
F_2=&&2xF_1\\ F_3=&&2\sum_i(g_V g_A)_i (q_i+\bar{q}_i)
\end{eqnarray}
where $(g_V)_i$, $(g_A)_i$ and $q_i$ are the weak charges and quark
distribution functions of the ith quark flavor. In our calculations
parton distribution functions of Martin, Roberts, Stirling and Thorne
(MRST2004) \cite{Martin:2004dh} have been used. We assume an isoscalar
nucleus $N=(p+n)/2$. This would be good assumption if the glass target
is pure $Si O_2$. Natural silicon is 92.2\% $^{28}$Si, 4.7\% $^{29}$Si,
and 3.1\% $^{30}$Si, where only $^{29}$Si is not isoscalar \cite{ti}.
Naturally occurring oxygen is 99.8\% $^{16}$O. Hence the error incurred
by assuming an isoscalar target would be not more than a few percent.

Possible new physics contributions coming from the operators
in (6) and (7)
only modify the lepton tensor:
\begin{eqnarray}
L_{\mu\nu}=\frac{1}{2}\left(1+\frac{v^2}{\Lambda^2}(
-\alpha_\mu+\alpha_\mu^\prime)\right)^2
\left(k_\mu k_\nu^\prime+k_\mu^\prime k_\nu-k\cdot k^\prime
g_{\mu\nu}+i\epsilon_{\mu\nu\alpha\beta}k^\alpha k'^\beta \right)
\end{eqnarray}
where, $k_\mu$ and $k_\mu^\prime$ are the momenta of initial and
final state neutrinos.
Therefore $\nu_\mu$ neutral current deep inelastic scattering
isolates the couplings $\alpha_\mu$ and $\alpha_\mu^\prime$. It does
not receive any contribution from $\alpha_e$ and $\alpha_e^\prime$.
As we have discussed this is not the case in $\nu_\mu e^- \to
\nu_\mu e^-$.

\begin{figure}
\includegraphics{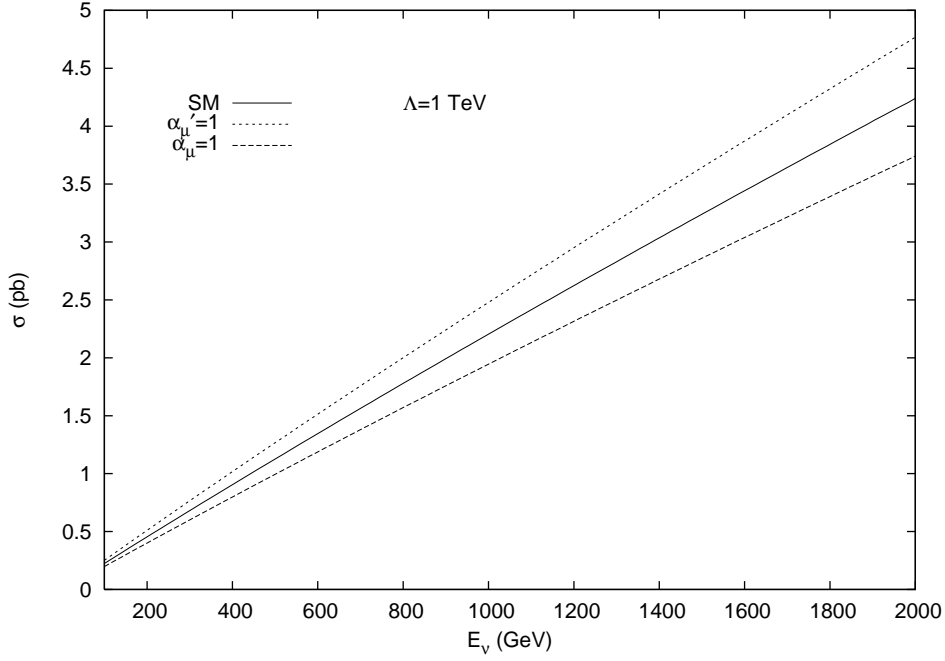}
\caption{Integrated total cross section of $\nu_\mu N \to \nu_\mu X$
as a function of initial neutrino energy for various values of the
anomalous couplings. Only one of the anomalous couplings is kept
different from their SM value. \label{fig5}}
\end{figure}

The behavior of the integrated total cross section as a function of
initial neutrino energy is plotted for various values of anomalous
couplings in Fig. \ref{fig5}. We see from the figure that cross
section has a linear energy dependence in the energy interval
$100-2000$ GeV. Deviation of the anomalous cross sections from their
SM value increase as the energy increases. Therefore high energy
neutrino experiments are  expected to reach a high sensitivity to
probe these anomalous couplings.

\begin{figure}
\includegraphics{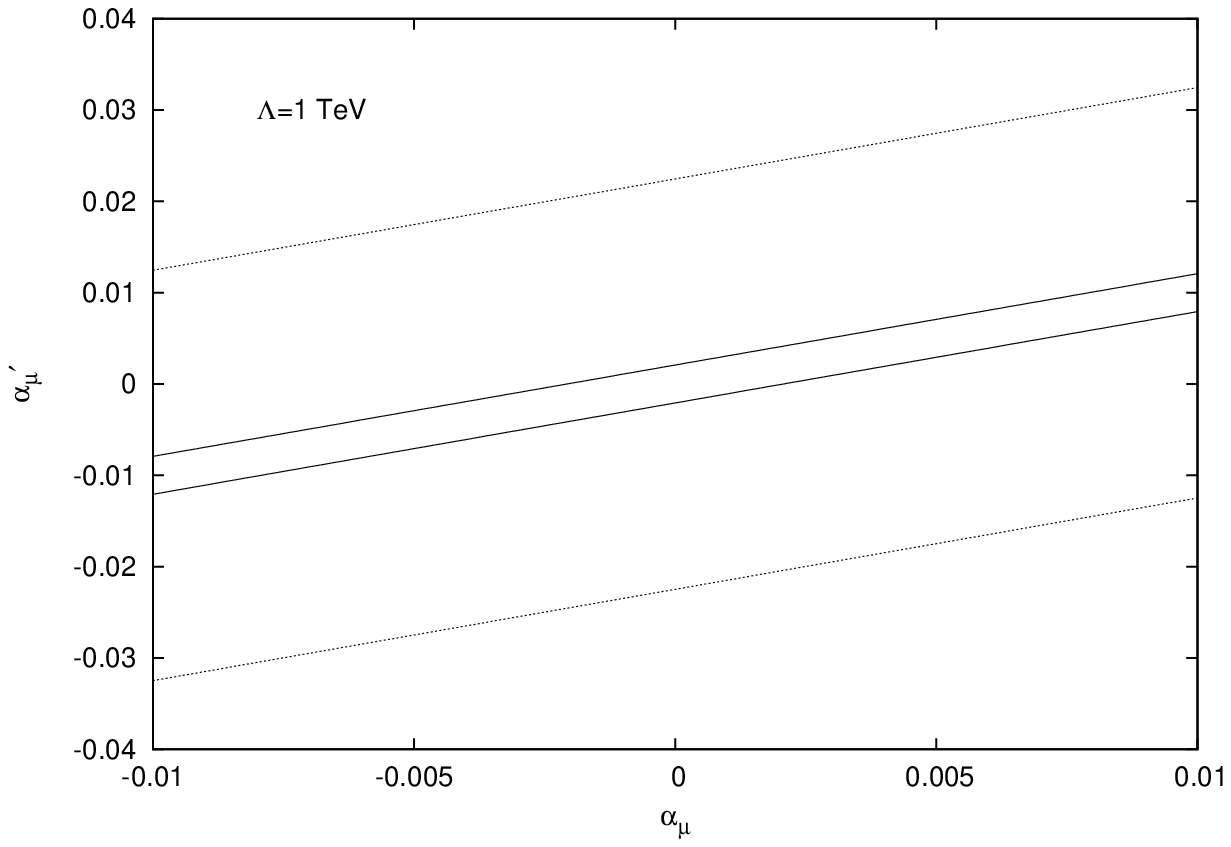}
\caption{The area bounded by the solid (dotted) lines is $95\%$ C.L.
sensitivity bound on the parameter space $\alpha_\mu -
\alpha_\mu^\prime$ for NuSOnG (NuTeV) statistics of $\nu_\mu N \to
\nu_\mu X$ scattering. \label{fig6}}
\end{figure}

In Fig.\ref{fig6} we show $95\%$ C.L. sensitivity bounds on the
parameter space $\alpha_\mu - \alpha_\mu^\prime$ for NuSOnG and
NuTeV statistics. We observe from the figure that NuSOnG has
approximately 10 times more sensitive bounds than NuTeV for $\nu_\mu
N \to \nu_\mu X$ scattering. Neutral current deep inelastic
scattering limits can be combined with $\nu_\mu e^- \to \nu_\mu e^-$
limits to place bounds on universality violating parameters
$\delta_1$ and $\delta_2$. Combining results of Fig. \ref{fig6} and
Fig. \ref{fig3} we obtain the bound $-0.074\leq \delta_2 \leq 0.074$
$(\alpha_\mu=0)$. Similarly combining Fig. \ref{fig6} and Fig.
\ref{fig2} we obtain the bound $-0.071\leq \delta_1 \leq 0.071$
$(\alpha_\mu^\prime=0)$. These bounds can be compared with CHARM
limits. From the first inequality of (5) we obtain $-2.2\leq\delta_2
\leq 3.3$ for $\delta_1=0$ and $-3.3\leq\delta_1 \leq 2.2$ for
$\delta_2=0$. Therefore  $\nu_\mu e^- \to \nu_\mu e^-$ and $\nu_\mu
N \to \nu_\mu X$ scattering processes at NuSOnG provide
approximately 40 times more restricted limits for $\delta_2$ and
$\delta_1$ compared with CHARM limits.

\section{Conclusions}

In some schemes new physics neutrinos participate may be
observable at lower energies, such as neutrino scattering through
an unparticle exchange \cite{Balantekin:2007eg}. However, to probe most
of the neutrino interactions beyond the Standard Model would require
energetic neutrino beams such as those employed in the NuSOnG proposal
or beta-beam proposals \cite{Mezzetto:2003ub,Volpe:2006in}. In this paper
we explored signatures for deviation from flavor universality in
neutrino-Z boson couplings. We found that the proposed  NuSOnG experiment
can place approximately an order of magnitude better limits than the CHARM
experiment in the muon-neutrino electron scattering mode.
We have also shown that deep-inelastic scattering measurements
with NuSOnG can place an almost two orders of magnitude better limits
on universality breaking than previous measurements.
Thus coupled with
possible complementary measurements of electron neutrino-electron
scattering cross section at beta beam experiments
\cite{Balantekin:2005md,de Gouvea:2006cb} NuSOnG experiment can be
a powerful probe of new neutrino physics.

\begin{acknowledgments}
We thank J. Conrad for discussions.
This work was supported in part by the
U.S. National Science Foundation Grant No.\ PHY-0555231 and in part by
the University of Wisconsin Research Committee with funds granted by
the Wisconsin Alumni Research Foundation.
\.{I}. \c{S}ahin and B. \c{S}ahin acknowledge
support through the Scientific and Technical Research Council
(TUBITAK) BIDEB-2219 grant.
\end{acknowledgments}

\pagebreak

\pagebreak

\begin{thebibliography}{99}

\bibitem{Yao:2006px}
  W.~M.~Yao {\it et al.}  [Particle Data Group],
  %``Review of particle physics,''
  J.\ Phys.\ G {\bf 33}, 1 (2006).
  %%CITATION = JPHGB,G33,1;%%

\bibitem{Masso:2002vj}
  E.~Masso,
  %``On universality of the coupling of neutrinos to Z,''
  Phys.\ Rev.\  D {\bf 66}, 077301 (2002)
  [arXiv:hep-ph/0205335].
  %%CITATION = PHRVA,D66,077301;%%

\bibitem{Vilain:1994gx}
  P.~Vilain {\it et al.}  [CHARM-II Collaboration],
  %``Constraints on additional Z bosons derived from neutrino - electron
  %scattering measurements,''
  Phys.\ Lett.\  B {\bf 332}, 465 (1994);
  %%CITATION = PHLTA,B332,465;%%
  %``Flavor Universality Of Neutrino Couplings With The Z,''
  Phys.\ Lett.\  B {\bf 320}, 203 (1994).
  %%CITATION = PHLTA,B320,203;%%

\bibitem{Dorenbosch:1986tb}
  J.~Dorenbosch {\it et al.}  [CHARM Collaboration],
  %``EXPERIMENTAL VERIFICATION OF THE UNIVERSALITY OF ELECTRON-NEUTRINO AND
  %MUON-NEUTRINO COUPLING TO THE NEUTRAL WEAK CURRENT,''
  Phys.\ Lett.\  B {\bf 180}, 303 (1986).
  %%CITATION = PHLTA,B180,303;%%

\bibitem{Adams:2008cm}
  T.~Adams {\it et al.}  [NuSOnG Collaboration],
  %``Terascale Physics Opportunities at a High Statistics, High Energy Neutrino
  %Scattering Experiment: NuSOnG,''
  arXiv:0803.0354 [hep-ph].
  %%CITATION = ARXIV:0803.0354;%%

\bibitem{Davidson:2003ha}
  S.~Davidson, C.~Pena-Garay, N.~Rius and A.~Santamaria,
  %``Present and future bounds on non-standard neutrino interactions,''
  JHEP {\bf 0303}, 011 (2003)
  [arXiv:hep-ph/0302093].
  %%CITATION = JHEPA,0303,011;%%

\bibitem{Bell:2005kz}
  N.~F.~Bell, V.~Cirigliano, M.~J.~Ramsey-Musolf, P.~Vogel and M.~B.~Wise,
  %``How magnetic is the Dirac neutrino?,''
  Phys.\ Rev.\ Lett.\  {\bf 95}, 151802 (2005)
  [arXiv:hep-ph/0504134].
  %%CITATION = PRLTA,95,151802;%%

\bibitem{Mohapatra:2005wg}
  R.~N.~Mohapatra {\it et al.},
  %``Theory of neutrinos: A white paper,''
  Rept.\ Prog.\ Phys.\  {\bf 70}, 1757 (2007)
  [arXiv:hep-ph/0510213].
  %%CITATION = RPPHA,70,1757;%%

\bibitem{Mohapatra:2006gs}
  R.~N.~Mohapatra and A.~Y.~Smirnov,
  %``Neutrino mass and new physics,''
  Ann.\ Rev.\ Nucl.\ Part.\ Sci.\  {\bf 56}, 569 (2006)
  [arXiv:hep-ph/0603118].
  %%CITATION = ARNUA,56,569;%%

\bibitem{de Gouvea:2007xp}
  A.~de Gouvea and J.~Jenkins,
  %``A Survey of Lepton Number Violation Via Effective Operators,''
  Phys.\ Rev.\  D {\bf 77}, 013008 (2008)
  [arXiv:0708.1344 [hep-ph]].
  %%CITATION = PHRVA,D77,013008;%%

\bibitem{Perez:2008ha}
  P.~Fileviez Perez, T.~Han, G.~y.~Huang, T.~Li and K.~Wang,
  %``Neutrino Masses and the LHC: Testing Type II Seesaw,''
  arXiv:0805.3536 [hep-ph].
  %%CITATION = ARXIV:0805.3536;%%





\bibitem{Buchmuller:1985jz}
  W.~Buchmuller and D.~Wyler,
  %``Effective Lagrangian Analysis Of New Interactions And Flavor
  %Conservation,''
  Nucl.\ Phys.\  B {\bf 268}, 621 (1986).
  %%CITATION = NUPHA,B268,621;%%

\bibitem{Marciano:2003eq}
  W.~J.~Marciano and Z.~Parsa,
  %``Neutrino-Electron Scattering Theory,''
  J.\ Phys.\ G {\bf 29}, 2629 (2003)
  [arXiv:hep-ph/0403168].
  %%CITATION = JPHGB,G29,2629;%%

\bibitem{Marciano:1987pf}
  W.~J.~Marciano and Z.~Parsa,
  %``ELECTROWEAK TESTS OF THE STANDARD MODEL,''
  Ann.\ Rev.\ Nucl.\ Part.\ Sci.\  {\bf 36}, 171 (1986).
  %%CITATION = ARNUA,36,171;%%

\bibitem{Kyuldjiev:1984kz}
  A.~V.~Kyuldjiev,
  %``Searching For Effects Of Neutrino Magnetic Moments At Reactors And
  %Accelerators,''
  Nucl.\ Phys.\  B {\bf 243}, 387 (1984).
  %%CITATION = NUPHA,B243,387;%%

\bibitem{Vogel:1989iv}
  P.~Vogel and J.~Engel,
  %``Neutrino Electromagnetic Form-Factors,''
  Phys.\ Rev.\  D {\bf 39}, 3378 (1989).
  %%CITATION = PHRVA,D39,3378;%%

\bibitem{Balantekin:2006sw}
  A.~B.~Balantekin,
  %``Neutrino magnetic moment,''
  AIP Conf.\ Proc.\  {\bf 847}, 128 (2006)
  [arXiv:hep-ph/0601113].
  %%CITATION = APCPC,847,128;%%

\bibitem{zeller} G. P. Zeller, "A precise measurement of the weak mixing
angle in neutrino nucleon scattering" (Thesis) UMI-30-50615.

\bibitem{Blumlein:1996vs}
  J.~Blumlein and N.~Kochelev,
  %``On the twist 2 and twist 3 contributions to the spin-dependent electroweak
  %structure functions,''
  Nucl.\ Phys.\  B {\bf 498}, 285 (1997)
  [arXiv:hep-ph/9612318].
  %%CITATION = NUPHA,B498,285;%%

\bibitem{Forte:2001ph}
  S.~Forte, M.~L.~Mangano and G.~Ridolfi,
  %``Polarized parton distributions from charged-current deep-inelastic
  %scattering and future neutrino factories,''
  Nucl.\ Phys.\  B {\bf 602}, 585 (2001)
  [arXiv:hep-ph/0101192].
  %%CITATION = NUPHA,B602,585;%%

\bibitem{Forte:2001zu}
  S.~Forte,
  %``Highlights of short baseline physics at a neutrino factory,''
  Nucl.\ Instrum.\ Meth.\  A {\bf 503}, 87 (2001)
  [arXiv:hep-ph/0109219].
  %%CITATION = NUIMA,A503,87;%%

\bibitem{Forte:2002ip}
  S.~Forte,
  %``Hadron physics at a neutrino factory,''
  Nucl.\ Phys.\  A {\bf 711}, 323 (2002)
  [arXiv:hep-ph/0207209].
  %%CITATION = NUPHA,A711,323;%%

\bibitem{Martin:2004dh}
  A.~D.~Martin, R.~G.~Roberts, W.~J.~Stirling and R.~S.~Thorne,
  %``Parton distributions incorporating QED contributions,''
  Eur.\ Phys.\ J.\  C {\bf 39}, 155 (2005)
  [arXiv:hep-ph/0411040].
  %%CITATION = EPHJA,C39,155;%%

\bibitem{ti}
{\em Table of Isotopes}, R.B. Firestone, V.S. Shirley, Eds. (Wiley,
New York, 1996).

\bibitem{Balantekin:2007eg}
  A.~B.~Balantekin and K.~O.~Ozansoy,
  %``Constraints on Unparticles from Low Energy Neutrino-Electron Scattering,''
  Phys.\ Rev.\  D {\bf 76}, 095014 (2007)
  [arXiv:0710.0028 [hep-ph]].
  %%CITATION = PHRVA,D76,095014;%%

\bibitem{Mezzetto:2003ub}
  M.~Mezzetto,
  %``Physics reach of the beta beam,''
  J.\ Phys.\ G {\bf 29}, 1771 (2003)
  [arXiv:hep-ex/0302007];
  %%CITATION = JPHGB,G29,1771;%%
  %``Physics potential of the gamma = 100,100 beta beam,''
  Nucl.\ Phys.\ Proc.\ Suppl.\  {\bf 155}, 214 (2006)
  [arXiv:hep-ex/0511005].
  %%CITATION = NUPHZ,155,214;%%

\bibitem{Volpe:2006in}
  C.~Volpe,
  %``Topical review on 'beta-beams',''
  J.\ Phys.\ G {\bf 34}, R1 (2007)
  [arXiv:hep-ph/0605033].
  %%CITATION = JPHGB,G34,R1;%%

\bibitem{Balantekin:2005md}
  A.~B.~Balantekin, J.~H.~de Jesus and C.~Volpe,
  %``Electroweak tests at beta-beams,''
  Phys.\ Lett.\  B {\bf 634}, 180 (2006)
  [arXiv:hep-ph/0512310].
  %%CITATION = PHLTA,B634,180;%%

\bibitem{de Gouvea:2006cb}
  A.~de Gouvea and J.~Jenkins,
  %``What can we learn from neutrino electron scattering?,''
  Phys.\ Rev.\  D {\bf 74}, 033004 (2006)
  [arXiv:hep-ph/0603036].
  %%CITATION = PHRVA,D74,033004;%%

\end{thebibliography}
\end{document}